\documentclass[twocolumn,nofootinbib,superscriptaddress,floatfix,noeprint,longbibliography]{revtex4-1}

\usepackage{graphicx,bm,units,tabularx,makecell}
\usepackage{amsmath,amssymb}
\usepackage[table]{xcolor}
\usepackage{hyperref}
\usepackage[nohug]{diagrams}
\diagramstyle[labelstyle=\scriptstyle]

\begin{document}

\title{Sources of quantized excitations via dichotomic topological cycles}

\author{Bryan Leung}
\email{bleung@spotify.com}
\affiliation{Spotify, New York, NY 10007, USA}

\author{Emil Prodan}
\email{prodan@yu.edu}
\affiliation{Department of Physics, Yeshiva University, New York, NY 10033, USA}

\begin{abstract}
We demonstrate the existence of a conceptually distinct topological pumping phenomenon in one-dimensional chains undergoing topological adiabatic cycles. Specifically, for a stack of two semi-infinite chains cycled in opposite directions and coupled at one edge by a gapping potential, we derive a higher-order bulk-boundary correspondence that relates the bulk Chern number associated with the adiabatic cycle of a single infinite chain and the number of electrons transferred between the semi-infinite chains. The relation is formulated using the relative index of two projections and proven using K-theoretic calculations. The phenomenon is exemplified using the Rice-Mele model and possible experimental implementations with classical and quantum degrees of freedom are discussed.
\end{abstract}

\maketitle
\section{Introduction}

Forty years ago, Thouless predicted that topological quantized pumping of charge can occur under an adiabatic cycle in a one-dimensional insulating bulk system with uniformly filled valence bands \cite{Thouless1983}. He arrived at this conclusion via a bulk calculation involving the physical observables and the states of the infinite system. Thouless never considered a physical boundary, because his calculation was all about the flow of charge through a theoretical section of the infinite system. Nevertheless, Thouless pumping is often presented as the quantum analog of the Archimedean screw (see {\it e.g.} Ref.~\cite{AltshulerScience1999}), which loads water at one end and spills water at the other end every time the screw is cranked up. For this to happen at the quantum level, it is often suggested that a half-infinite Thouless chain needs to be put in contact with a metal, though the precise pheonomenology of such an experiment is impossible to formalize by a clean calculation as in Ref.~\cite{Thouless1983}, because one is now dealing with a hybrid ungapped system. In this work, we present an experimentally verifiable scenario in which topological quantized pumping is achieved at the contact between a Thouless chain and an empty and spectrally gapped system. This is interesting because the pumped electrons or the excitations are not mixed with the Fermi sea of a metal. For this reason, we claim that our proposed mechanism supplies the principle for a genuine source of quantized excitations. Furthermore, the effect can be observed in phonic and photonic crystals as well.

There are two types of pumpings observed in experiments recently. The first type involves Thouless quantized pumping in the bulk of one-dimensional systems with ultracold fermionic and bosonic atoms \cite{NakajimaNature2016,LohseNature2016}. The second type is the edge-to-edge pumping which relies purely on the spectral flow and on the adiabatic theorem \cite{GrinbergNatComm2020,XiaPRL2021,ChengPRL2020,XuPRL2020}. In such experiments, the spectral bands are empty and a localized mode is loaded at one end of a chain. Upon an adiabatic deformation of the system, the mode follows the spectral flow and ends up at the other side of the finite system. This type of pumping relies on the bulk-boundary correspondence principle which warrants a chiral spectral flow, and this is all needed for the success of such experiments. We mention this type of experiments because the setting is that of a finite chain in contact with a vacuum, but it will also work if the contact is with an insulator. However, the physical processes are very different from those involved in Thouless pumping. Specifically, there are no dynamics of bulk states because the spectral bands are empty at the start. A topological pumping phenomenon similar to the one presented here was discovered in Ref.~\cite{LeungJPA2020}, in the context of three-dimensional condensed matter systems displaying the quantized magnetoelectric effect. In this work, we demonstrate that this topological pumping effect also occurs in weakly coupled one-dimensional chains under topological adiabatic cycles. Since these systems are much easier to handle in a laboratory, we hope that our present theoretical work will eventually lead to an experimental realization.

\begin{figure}[!b]
\center
\includegraphics[width=1\linewidth]{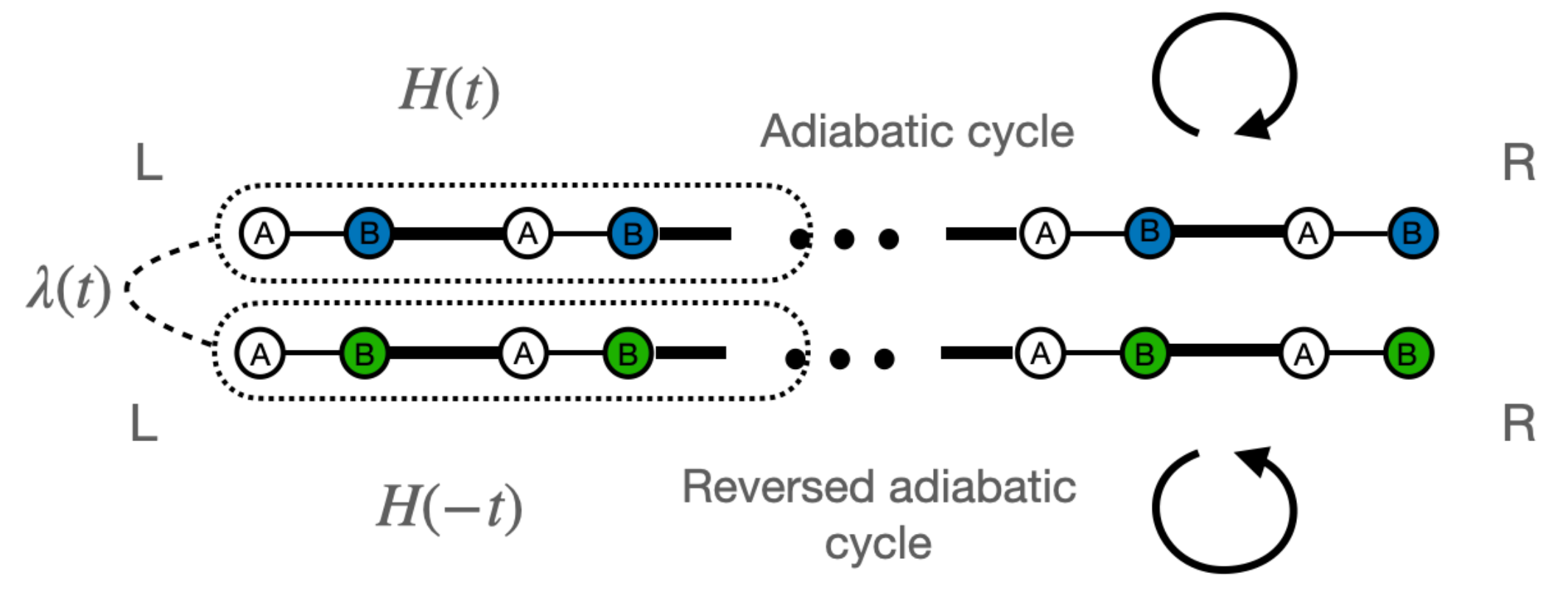}
\caption{\small Illustration of our coupled Rice-Mele chains adiabatically driven in opposite topological cycles, with the two sublattice sites $A$ and $B$ of unit cells labeled. At $t=0$, the valence bands of the top chain (blue) are uniformly filled and the energy bands of the bottom chain (green) are completely empty. The coupling $\lambda (t)$ switches on the edge potential $\bm V_{\rm edge}(t)$, defined by Eq.~\eqref{Eq:EdgeV}, which couples the chiral edge bands of the top and bottom chains.}
\label{fig:Fig1}
\end{figure}

\begin{figure*}[t]
\center
\includegraphics[width=1\linewidth]{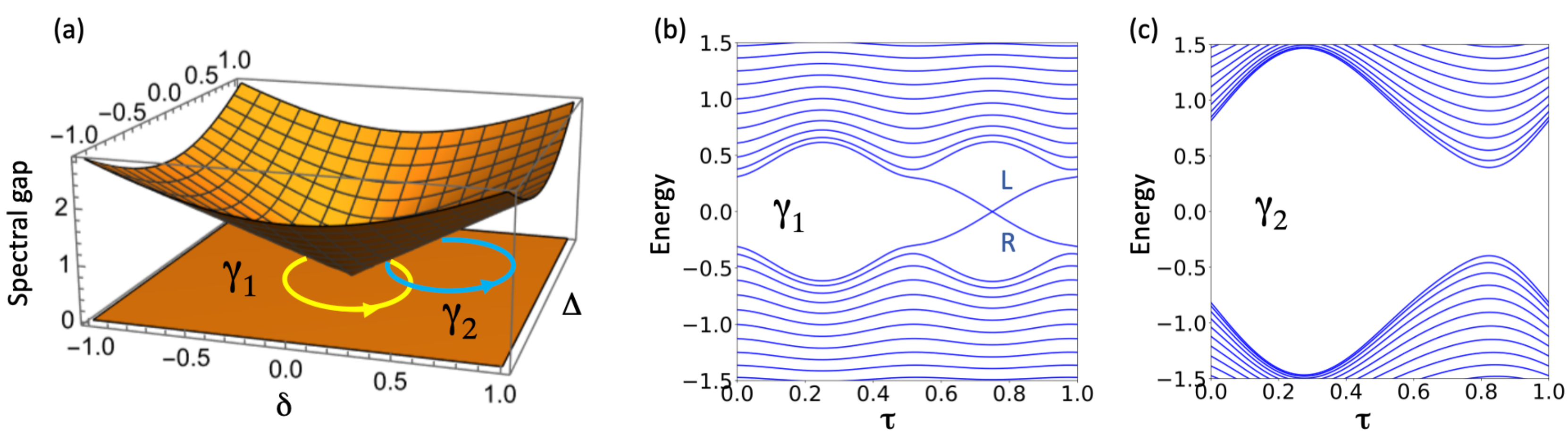}
\caption{\small (a) Spectral gap of the Hamiltonian \eqref{Eq:RM_Ham} as a function of the parameters $(\delta,\Delta)$, together with two adiabatic cycles, of which $\gamma_1 = (\delta=0.3 \sin(2\pi \tau), \Delta=-0.3\cos(2\pi \tau) )$ circles and $\gamma_2= (\delta=0.4+0.3 \sin(2\pi \tau),\Delta=0.4-0.3\cos(2\pi \tau))$ does not circle the gap singularity at $\delta=\Delta=0$. (b,c) The energy spectrum of the Hamiltonian \eqref{Eq:RM_Ham} as a function of the adiabatic parameter $\tau$ for $\gamma_1$ and $\gamma_2$, respectively.  Both spectra are generated with chains of 20 sites. Two chiral modes located at opposite left (L) and right (R) ends of the chain are visible in (b). These chiral modes are absent for the adiabatic cycle $\gamma_2$.}
\label{fig:Fig2}
\end{figure*}

\section{Setting and main result}

The proposed setup, illustrated in Fig.~\ref{fig:Fig1}, consists of a stack of two semi-infinite Rice-Mele chains \cite{RiceMele1982} driven in opposite topological adiabatic cycles. By unfolding, the system can also be thought of as two interfaced half-infinite Rice-Mele chains driven by identical adiabatic cycles. We, however, will work with the folded configuration which transforms an interface problem into a bulk-edge problem, for which there exist many tools of analysis. Now, at $t=0$, the valence bands of the top chain are uniformly filled and all energy bands of the bottom chain are completely empty, and the two chains are decoupled. As the adiabatic cycles progress, the coupling between the chains is turned on and then off towards the end of the cycles. For this hybrid system, suppose $\widehat{\bm  \Pi}_0$ is the projection onto the populated states at $t=0$ and $\widehat{\bm \Pi}^A_1$ is the adiabatic time evolution of this projection after a complete cycle. We show that
\begin{equation}\label{Eq:RelativeIndChern0}
    {\rm Index}\big( \widehat{\bm \Pi}^A_1, \ \widehat{\bm  \Pi}^{}_0 \big) = {\rm Ch}_\gamma,
\end{equation}
where on the left we have the relative index of Avron, Seiler and Simon \cite{AvronPRL1990,AvronCMP1994,AvronJFA1994} and on the right is the Chern number~\eqref{Eq:Chern} associated with the adiabatic cycle of a single infinite chain. This identity connects a two-dimensional topological invariant and a zero-dimensional invariant related to the interface, hence it is a higher-order bulk-boundary correspondence. It tells us that a ${\rm Ch}_\gamma$ number of electrons have been pumped into the system and, since the top chain had all the available states occupied, these extra electrons must have been pumped into the bottom chain. Note that a relation similar to Eq.~\eqref{Eq:RelativeIndChern0} is needed to rigorously justify the analogy mentioned at the beginning between pumping into a metal and the Archimedean screw, but no such relation has been derived. This shows how poor our understanding of Thouless pumping in the presence of physical interfaces is, and hopefully it clarifies the importance of the step taken by our work.

\section{Model, insights and experimental implementations}

Let us consider a single Rice-Mele Hamiltonian \cite{RiceMele1982}
\begin{align}\label{Eq:RM_Ham}
    H =& - \sum_{x} \big (J_{1}|x, B \rangle \langle x, A| + J_{2}|x+1, A \rangle \langle x, B| + h.c. \big) \\ \nonumber 
    &+  \Delta \sum_{x} \big( |x, A \rangle \langle x, A| - |x, B \rangle \langle x, B|\big ),
\end{align}
where $A$ and $B$ denote the sublattice sites of the $x$-th unit cell, and $J_{1/2} = 1 \pm \delta$ and $\Delta$ denote the tunneling couplings and energy offset between neighboring sites, respectively. Figure~\ref{fig:Fig2}(a) shows its spectral gap mapped as a function of the parameters $\delta$ and $\Delta$, together with two adiabatic paths parametrized by the circle $\mathbb S^1 = \mathbb R/\mathbb Z$, of which $\gamma_1$ encloses the gap singularity while $\gamma_2$ does not. The Chern number for a closed and spectrally gapped loop $\gamma$ is \cite{NiuRMP2010}
\begin{equation}\label{Eq:Chern}
{\rm Ch}_\gamma = {\rm i} \int_{\mathbb S^1} d\tau \; {\rm Tr}_L \big ( P(\tau) \big [ \partial_\tau  P(\tau),\mathrm{i} [X, P(\tau)] \big ]  \big ),
\end{equation}
where $P(\tau)=\chi_{(-\infty,E_F]} ( H(\gamma_\tau) )$ is the spectral projector onto the lower energy bands below the gap, $X$ is the position operator and ${\rm Tr}_L$ is the trace per length. Throughout, $\chi_A $ denotes the characteristic function of the set $A$. ${\rm Ch}_\gamma$ is written in real space rather than $k$-space to convey that it is well defined in the presence of disorder \cite{BellissardJMP1994}. The difference between $\gamma_1$ and $\gamma_2$ is that ${\rm Ch}_{\gamma_1}=1$ while ${\rm Ch}_{\gamma_2}=0$. The expected chiral edge modes traversing the spectral bulk gap for $\gamma_1$ are illustrated in Fig.~\ref{fig:Fig2}(b). 

\begin{figure*}[t]
\center
\includegraphics[width=1\linewidth]{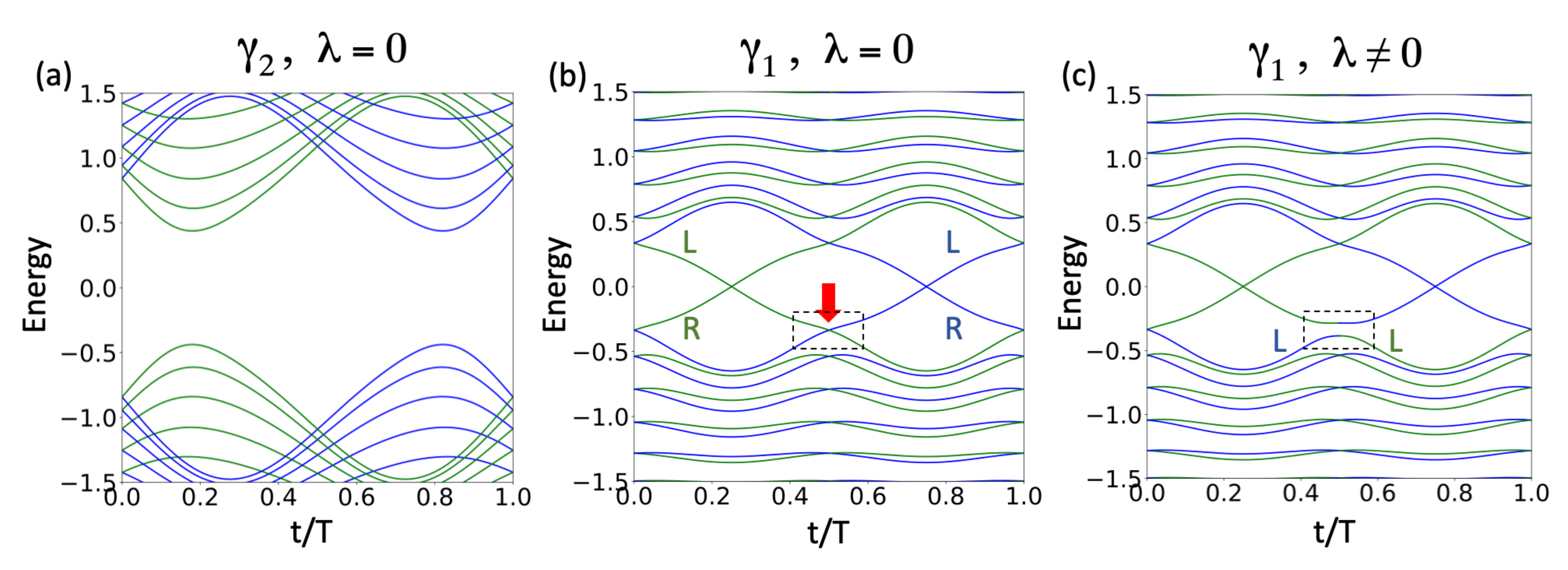}
\caption{\small Spectral flow of the Hamiltonian~\eqref{Eq:UncoupledHam} reduced on finite-size chains with open boundary conditions, when adiabatically deformed along (a) the cycle $\gamma_2$ and (b) the cycle $\gamma_1$. (c) Same as (b) but for the Hamiltonian~\eqref{Eq:FullHam}. The labels L and R indicate the spatial localization of the chiral bands to either the left (L) or the right (R) edge of the system. Marked with an arrow in (b) is the crossing of the chiral bands localized at the left edge of the system. The window drawn with a dashed line indicates the energy and time ranges where the edge potential~\eqref{Eq:EdgeV} is defined. The data are generated with finite chains containing 10 sites and $\lambda(t)$ for (c) is turned up to 0.05. The blue and green colors represent the overlap with the top and bottom states respectively.}
\label{fig:Fig3}
\end{figure*}

We now consider a stack of two infinite Rice-Mele chains and an adiabatic cycle $\gamma$, but we run the adiabatic cycle in opposite directions for the two blocks, namely
\begin{equation}\label{Eq:UncoupledHam}
\bm H_0(t) = \begin{pmatrix} H(\gamma_{\tau}) & 0 \\ 0 & H(\gamma_{-\tau}) \end{pmatrix}, \quad \tau=\frac{t}{T}, 
\end{equation}
where $t$ is the real time and $T$ is the total duration of cycle. It is convenient to use $T$ as the unit of time, in which case $t$ and $\tau$ coincide and we can use the latter throughout. Hamiltonian $\bm H_0(\tau)$ plays the role of the bulk Hamiltonian. Throughout, our convention is to bold all operators related to the double-chain system. Now, if one calculates the total Chern number of the lower energy bands, one will find a trivial value, regardless if $\gamma$ encloses the critical point or not. As such, one will be inclined to proclaim that all topological characteristics are lost. Our main message is that this is not the case at all! Let us examine the spectral flows reported in Figs.~\ref{fig:Fig3}(a) and \ref{fig:Fig3}(b) of the Hamiltonian~\eqref{Eq:UncoupledHam}, when it is reduced to a finite-size with open boundary conditions. The results in Fig.~\ref{fig:Fig3}(a) for the contour $\gamma_2$ are not very interesting, but an opportunity presents itself for contour $\gamma_1$ in Fig.~\ref{fig:Fig3}(b). Indeed, two opposite chiral bands spatially localized at the {\it same} edge intersect each other as indicated by the arrow. Then a generic edge potential hybridizes these bands and gaps the system at that edge, potentially resulting in a quantized spectral flow from the top to the bottom chain. Let us point out that Eq.~\eqref{Eq:RelativeIndChern0} prohibits the emergence of other band splittings undoing the effect.

To exemplify that the effect is possible, we design an edge potential that hybridizes only the two chiral bands from the left edges, leaving the rest of the spectrum {\it intact}. This irrefutably proves that the transfer of quantized excitations from one chain to the other is possible. The topological character of the process, that is, its robustness against deformations of the bulk models and edge potentials as well as the inclusion of disorder, is a separate question. This is addressed in the next section. In the following computations, all Hamiltonians are assumed finite and with open boundary conditions.  Now, let us consider an energy window $\Delta E=[E_-, E_+ ]$ around the band crossing (see Fig.~\ref{fig:Fig3}(b)) and let $P_{\Delta E}(\tau) = \chi_{[E_-, E_+ ]}( H (\gamma_\tau) )$ be the corresponding spectral projection of the top chain. Then our proposed edge potential takes the form 
\begin{equation}\label{Eq:EdgeV}
\bm V_{\rm edge} (\tau) = \lambda (\tau) \begin{pmatrix} 0 & P_{\Delta E} (\tau) P_{\Delta E} (-\tau) \\ P_{\Delta E} (-\tau) P_{\Delta E} (\tau) & 0 \end{pmatrix},
\end{equation}
where $\lambda(\tau)$ is a smooth on-off switch that is zero outside the small window shown in Fig.~\ref{fig:Fig3}(b) centered at the crossing point. Note that $P_{\Delta E}(\tau)$ displays discontinuities when the chiral bands cross the edges of the interval $\Delta E$, but those discontinuities disappear once the smooth on-off switch is included. Moreover, the spectral projection of $\bm H_0(\tau)$ on the interval $\Delta E =[E_-, E_+ ]$,
\begin{equation} 
\bm Q_{\Delta E}(\tau) := \chi_{[E_-, E_+ ]}\big ( \bm H_0(\tau) \big)= \begin{pmatrix} P_{\Delta E}(\tau) & 0 \\ 0 & P_{\Delta E}(-\tau) \end{pmatrix},
\end{equation} 
commutes with our specially designed edge potential and this assures us that only the states inside the spectral interval $[E_-,E_+]$ are affected by $\bm V_{\rm edge}(\tau)$.

The spectral flow of the full Hamiltonian
\begin{equation}\label{Eq:FullHam}
    \bm H(\tau) = \bm H_0(\tau) + \bm V_{\rm edge}(\tau), \quad \tau = t/T,
\end{equation}
illustrated in Fig.~\ref{fig:Fig1}, is depicted in Fig.~\ref{fig:Fig3}(c). As expected, the spectrum has been modified only inside the marked window where one can clearly distinguish a splitting of the left-localized chiral bands. Outside the marked window, the two Rice-Mele chains are decoupled, and hence the eigenstates have a well defined top and bottom index, which is color labeled in Fig.~\ref{fig:Fig3}. Based on this concrete spectral flow, the following statement holds: If at $t=0$ one fully populates the states below the bulk spectral gap of the top chain and leaves the states of the bottom chain completely empty, then exactly one normalized state will be detected on the bottom chain after one full adiabatic cycle.

It is now clear that the adiabatic cycle described above acts like a valve that releases one excitation per cycle into the bottom chain. Hence, our proposed setup supplies the design principle for on-demand sources of quantized excitations. Here are two possible laboratory implementations. The first one involves phonons. In this context, the sites of the Rice-Mele chains harbor identical mechanical resonators which are coupled as indicated by our Hamiltonian \eqref{Eq:FullHam}. This requires fast dynamic re-configurability to a level where the pumped signals do not succumb to dissipation and the effect can be revealed. The edge-to-edge pumping experiments mentioned in the beginning have demonstrated that this is now under control in metamaterials. With the chains decoupled, the upper chain is excited with a broad spectrum source such that all phonon modes below the spectral gap are uniformly excited. After one turn of our proposed cycle, a {\it single} phonon will be detected propagating to the right on the lower mechanical chain. The situation is more difficult with electrons because populating and de-populating energy bands is a more involved process. Still, we believe that the effect can be implemented with a half-filled virtual spin-Chern insulator. By applying a strong upward magnetic field, we can populate the up-spin states and completely depopulate the down-spin states. The magnetic field can then be abruptly turned off and the adiabatic cycle can be initiated. After one turn of the adiabatic cycle, one should observe a quantized spin-down excitation moving to the right. Of course, all these will take place in the background of the relaxation process back to the equilibrium state. Hence, the success of such an enterprise requires the adiabatic cycle to be shorter than the relaxation time. The recent cold-atom simulations of spin-Chern insulators \cite{LvPRL2021} show a level of control that we believe is sufficient for implementing and observing the effect proposed here.

\section{Rigorous analysis}
We now address the topological character of the process and show that the phenomenon is independent of the particular form of the coupling potential. To derive the bulk-edge correspondence, we send the right edge to $\infty$, and hence the adiabatically cycled system is now semi-infinite. By doing so, we clear up the spectral gap of the right chiral bands and, as such, we can apply the adiabatic theorem to the lower spectrum. To distinguish between the bulk and the half-space operators, we will place a hat on the latter. Hence, the Hamiltonian~\eqref{Eq:FullHam} becomes $\widehat{\bm H} (\tau)$. Furthermore, the operators with matrix elements decaying to zero away from the edge will carry a tilde. Hence, the potential~\eqref{Eq:EdgeV} becomes $\widetilde{\bm V}_{\rm edge} (\tau)$. If $\widehat{\bm U}_{\tau}$ is the physical time evolution operator, {\it i.e.} the unitary solution of the equation, ${\rm i} \partial_\tau \widehat{\bm U}_{\tau} = T\, \widehat{\bm H}(\tau)\widehat{\bm U}_{\tau}$, $\widehat{\bm U}_{0}=\bm 1$, then it is well known \cite{AvronCMP1987} that $\widehat{\bm U}_{\tau}$ can be approximated by the adiabatic time evolution $\widehat{\bm U}^A_{\tau}$ (see Supplemental Material \cite{Supplemental}). Then our task is to compare the projections
\begin{equation}\label{Eq:Proj}
    \widehat{\bm \Pi}^{}_0=\begin{pmatrix} \widehat P(0) & 0 \\ 0 & 0 \end{pmatrix} \ \ {\rm and} \ \   \widehat{\bm \Pi}^A_{1}=\widehat{\bm U}^A_{1}\begin{pmatrix} \widehat P(0) & 0 \\ 0 & 0 \end{pmatrix}\widehat{\bm U}_{1}^{A\ast},
\end{equation}
encoding the initial and final states of the two-chain system for large $T$. Here, $\widehat P(0)=\chi_{(-\infty,E_F]} ( \widehat H(\gamma_0) )$ is the spectral projection of the top chain at the beginning of the cycle and $E_F$ lies in the spectral gap of $\widehat H(\gamma_{\tau})$.

Let us point out again the unusual character of the setting. On one hand, the bulk topology, prompted by the non-zero Chern number ${\rm Ch}_{\gamma_1}$, exists in the $(1+1)$-dimensional space (one real and one  virtual dimension). On the other hand, Eq.~\eqref{Eq:Proj} compares projections on a semi-infinite quasi-one dimensional system because time does not appear as a variable in Eq.~\eqref{Eq:Proj}. As we shall see, this comparison provides a topological invariant that is associated with the edge physics of the semi-infinite static system. Hence, our task is to connect the bulk topology of a $(1+1)$-dimensional system with the edge topology of a quasi one-dimensional system, and this is a jump from a two-dimensional system to a zero-dimensional one. Thus, we are dealing with a higher-order topological phenomenon.

Now, the bulk Hamiltonians, such as $H$ or $\bm H_0$, belong to the algebra of periodic operators and are denoted by $\mathcal A$ in the following (adding or multiplying periodic operators preserves this property). The adiabatically driven Hamiltonians, such as $H({\gamma_\tau})$ or $\bm H_0(\tau)$, then belong to the algebra $\mathcal A^{\mathbb S^1}$ of maps from the circle $\mathbb S^1$ to the algebra $\mathcal A$. The operators with matrix elements concentrated around the edge live in the edge algebra $\widetilde{\mathcal A}$ (indeed, the sums and products of edge operators remain edge operators). In between $\mathcal A$ and $\widetilde{\mathcal A}$ is the algebra $\widehat{\mathcal A}$ of half-space operators whose elements are periodic half-infinite Hamiltonians with clean open boundary conditions plus any elements from the edge algebra, {\it i.e.} $\widehat {\bm H} = {\bm H}|_{\rm open}+\widetilde {\bm V}_{\rm edge}$. The projections from Eq.~\eqref{Eq:Proj} belong to $M_2 \otimes \widehat{\mathcal A}$, where $M_N$ denotes the algebra of $N\times N$ matrices. Two projections are said to be homotopic if they can be deformed continuously into each other without leaving the algebra they belong to. Since in condensed matter physics and metamaterial science we always work with effective models, it is more appropriate to include additional trivial bands, accounting for the neglected orbitals or degrees of freedom, and allow the deformations to spill out into these additional bands. In other words, to allow the deformations to take place in $M_{\infty} \otimes \widehat{\mathcal A}$, where $M_\infty$ is the algebra of arbitrary rank matrices. In this case, one talks about stable homotopy and the K-theoretic group $K_0(\widehat{\mathcal A})$ classifies the projections with respect to this equivalence relation. Note that the stable homotopy class $[\Pi]_0$ of a projection from $M_{\infty} \otimes \widehat{\mathcal A}$ is the {\it complete} topological invariant associated with $\Pi$ \cite{Footnote1}. All the above algebras can be trivially extended to include weak disorder, and more details on the algebras can be found in the Supplemental Material \cite{Supplemental}.

A K-theoretic calculation detailed in the Supplemental Material \cite{Supplemental} shows that the difference 
\begin{equation}\label{Eq:CTop}
    \big [ \widehat {\bm \Pi}^A_{1} \big]_0 \ - \ \big [ \widehat{\bm \Pi}^{}_0 \big ]_0
\end{equation}
actually lands in the K-theoretic group $K_0(\widetilde{\mathcal A})$ of the edge algebra. Among other things, this implies that the difference between the projections from Eq.~\eqref{Eq:Proj} belongs to the simple algebra $M_\infty$. It is an important detail because it enables us to connect to the work by Avron, Seiler and Simon on the relative index of projections \cite{AvronPRL1990,AvronCMP1994,AvronJFA1994}. Their original application was on a comparison between the Fermi projections of two-dimensional systems with and without Dirac fluxes piercing the plane of the sample. The mathematical concept supplied a rigorous framework for Laughlin's pumping argument for the integer quantum Hall effect \cite{Laughlin}. In our context, the relative index of the projections~\eqref{Eq:Proj} is a numerical topological invariant derived from the complete topological invariant~\eqref{Eq:CTop}, which measures their relative dimension.

We are now ready to state precisely our main result: For any edge potential $\widetilde {\bm V}_{\rm edge}(\tau)$ that gaps the spectrum for the entire duration of the cycle and vanishes for $\tau$ in small interval $[-\epsilon,\epsilon]$ around the initial and final points, we have the relative index
\begin{equation}\label{Eq:RelativeIndChern}
    {\rm Index}\big( \widehat{\bm \Pi}^A_1, \ \widehat{\bm  \Pi}^{}_0 \big) = {\rm Ch}_\gamma,
\end{equation}
where on the right is the Chern number~\eqref{Eq:Chern} associated with a single infinite chain. The above identity follows from a relation derived in Ref.~\cite{Supplemental} between the complete invariant~\eqref{Eq:CTop} and the complete topological invariant $[\tau \mapsto P(\tau)]_0 \in K_0 (\mathcal A^{\mathbb S^1} )$, which carries the Chern number ${\rm Ch}_\gamma$. It takes the form
\begin{equation}\label{Eq:TopIndexMap}
    \big [ \widehat{\bm \Pi}^A_{1} \big]_0 \ - \ \big [ \widehat{\bm \Pi}^{}_{0} \big ]_0 = \big( {\rm Ind}\circ \theta^{-1} \big)\big([\tau \mapsto P(\tau)]_0\big),
\end{equation}
where $\theta$ is the theta map appearing in the Bott periodicity theorem and ${\rm Ind}$ is the K-theoretic connecting map associated with the exact sequence of algebras $\widetilde{\mathcal A} \rightarrowtail \widehat{\mathcal A} \twoheadrightarrow \mathcal A$ \cite{Supplemental, RLL}. Small disorder is automatically included in these calculations.

In terms of the spectral flows depicted in Fig.~\ref{fig:Fig3}, we can interpret these predictions in the following way. Note that for $\gamma_1$ in Fig.~\ref{fig:Fig3}(c), there is an additional swap between an occupied and an unoccupied state for the top chain. But this swap takes place at the right edge and, when the right edge is sent to $\infty$, an observer operating at the origin will not be able to detect it. To this observer, it will appear that one state below the gap has been populated during the adiabatic cycle, without depopulating any other states. As such, this newly populated state must be residing on the bottom chain. In contrast, for $\gamma_2$, while an edge potential can still hybridize top and bottom non-chiral edge bands, there will be an equal number of top-bottom and bottom-up swaps of states, all taking place at the left edge. Hence, the observer will not detect any net increase in the population of the states. Furthermore, the nontopological edge bands are unstable and can disappear under deformations.

\section{Conclusion}

In conclusion, we have discovered a distinct pumping process which can lead to new applications of topology in condensed matter physics and metamaterial science. Based on our rigorous statements, we can assure the experimentists that no fine-tuning of the edge potential is required to achieve the quantized effect because a generic edge potential will typically gap the spectrum. Furthermore, the effect is robust against deformations and inherent small design imperfections.

\begin{acknowledgments}
E. P. acknowledges support from the National Science Foundation under Grants No. DMR-1823800 and No. CMMI-2131760.
\end{acknowledgments}

\clearpage

\appendix
\begin{center}
\textbf{Supplemental Material}
\end{center}

We provide the formalism and analysis supporting our statements in the main text.

\section{Adiabatic theorem and monodromies of projections}
\label{AppendixA}

The physical time evolution under the full Hamiltonian $\widehat{\bm H}(t)$ is generated by the time propagator $\widehat{\bm U}_{t}$, which is the unique unitary solution of the Schr\"{o}dinger equation
\begin{equation}
    {\rm i} \partial_t \widehat{\bm U}_{t} = \widehat{\bm H}(t)\widehat{\bm U}_{t}, \quad \widehat{\bm U}_{0}=\bm 1. 
\end{equation}
If one passes to the scaled time $\tau=t/T \in [ 0, 1 ]$ using the period $T$ of the cycle, the equation becomes
\begin{equation}
    {\rm i} \partial_\tau \widehat{\bm U}_{\tau} = T\,   \widehat{\bm H}(\gamma_\tau)\widehat{\bm U}_{\tau}, \quad \widehat{\bm U}_{0}=\bm 1. 
\end{equation}

The adiabatic time propagator $\widehat{\bm U}^A_{\tau}$, as defined in Ref.~\cite{AvronCMP1987} [Eq.~1], is the unique unitary solution of the following equation 
\begin{equation}\label{Eq:TimeEvol1}
{\rm i} \partial_\tau \widehat{\bm U}^A_{\tau} = \big (T \, \widehat{\bm H}(\gamma_\tau)+i [ \partial_\tau \widehat{\bm Q}(\tau), \widehat{\bm Q}(\tau) ]\big ) \widehat{\bm U}^A_{\tau},
\end{equation}
with the initial condition $\widehat{\bm U}^A_{0} = \bm 1$ and 
\begin{equation}
\widehat{\bm Q}(\tau)=\chi_{(-\infty,E_F]}\big(\widehat{\bm H}(\gamma_\tau)\big)
\end{equation}
the spectral projection of $\widehat{\bm H}(\gamma_\tau)$ onto the spectrum below its spectral gap. Note that the original work by Kato \cite{Kato} only considered the adiabatic evolution of a Hamiltonian that is completely degenerate on its projection. In this case, there exists a function ${\bm \lambda}(\gamma_\tau)$ such that $ \widehat{\bm H}(\gamma_\tau) \widehat{\bm Q}(\tau) = {\bm \lambda} (\gamma_\tau) \widehat{\bm Q}(\tau)$ for all $\tau$, and our adiabatic equation ~\eqref{Eq:TimeEvol1} can be expressed as 
\begin{equation}\label{Eq:DegenerateTimeEvol1}
{\rm i} \partial_\tau \widehat{\bm U}^A_{\tau} = \big (T {\bm \lambda} (\gamma_\tau) +i [ \partial_\tau \widehat{\bm Q}(\tau), \widehat{\bm Q}(\tau) ]\big ) \widehat{\bm U}^A_{\tau}.
\end{equation}
The $T {\bm \lambda} (\gamma_\tau)$ term can be removed with an appropriate choice of phase factor, then Eq.~\eqref{Eq:DegenerateTimeEvol1} reduces to the adiabatic equation
\begin{equation}\label{Eq:KatoEq}
{\rm i} \partial_\tau \widehat{\bm U}^A_{\tau} = i [ \partial_\tau \widehat{\bm Q}(\tau), \widehat{\bm Q}(\tau) ] \widehat{\bm U}^A_{\tau},
\end{equation}
commonly found in the literature \cite{Kato, AvronCMP1987}. However, we are dealing with a non-degenerate band, hence the appropriate adiabatic propagator is the one defined by Avron, Seiler and Yaffe \cite{AvronCMP1987}.

The adiabatic theorem states that \cite{AvronCMP1987}[p.~42]
\begin{equation}
    \widehat{\bm U}_\tau^\ast \, \widehat{\bm U}_\tau^A = I + {\mathcal O} (1/T) \quad \forall \, \tau \in [0,1].
\end{equation}
This, in turn, assures us that 
\begin{equation}
    \widehat{\bm U}_\tau \,  \widehat{\bm \Pi}^{}_0 \; \widehat{\bm U}_\tau^\ast - \widehat{\bm U}_{\tau}^A \, \widehat{\bm \Pi}^{}_0 \; \widehat{\bm U}_{\tau}^{A\ast} = {\mathcal O} (1/T),
\end{equation}
which justifies the use of $\widehat{\bm \Pi}_\tau^A$ in our statements. The hallmark property of $\widehat{\bm U}_{\tau}^A$ is that
\begin{equation}
\widehat{\bm U}_{\tau}^A \,   \widehat{\bm Q}(0) \, \widehat{\bm U}_{\tau}^{A \ast}=  \widehat{\bm Q}(\tau).
\end{equation}
As such,
\begin{equation}\label{Eq:FullMonodromy}
    \widehat{\bm W}_{\tau} : =  \widehat{\bm U}_{\tau}^A \,   \widehat{\bm Q}(0)
\end{equation}
supplies a monodromy of $\widehat{\bm Q}(\tau)$, {\it i.e.} a partial isometry with the properties
\begin{equation}
  \widehat{\bm W}_{\tau}^\ast \, \widehat{\bm W}_{\tau} =  \widehat{\bm Q}(0), \quad \widehat{\bm W}_{\tau} \, \widehat{\bm W}_{\tau}^\ast =  \widehat{\bm Q}(\tau).
\end{equation}
In particular, \begin{equation}
  \widehat{\bm W}_{1}^\ast \, \widehat{\bm W}_{1} =  \widehat{\bm Q}(0), \quad \widehat{\bm W}_{1} \, \widehat{\bm W}_{1}^\ast =  \widehat{\bm Q}(1)=\widehat{\bm Q}(0),
\end{equation}
and 
\begin{equation}\label{Eq:TheLift}
\widehat{\bm W}_{1}+\bm 1 - \widehat{\bm Q}(0)
\end{equation}
is a unitary operator from the algebra $\widehat{\mathcal A}$.

\section{Algebras of operators and their relations}
\label{AppendixB}

The algebras of physical operators $\mathcal A$,  $\widehat{\mathcal A}$, $\widetilde{\mathcal A}$ and ${\mathcal A}^{\mathbb S^1}$ introduced in the main text can be connected via exact sequences. One well-known short exact sequence of $C^\ast$-algebras is \cite{ProdanSpringer2016A}
\begin{equation}
\label{Eq:BulkBoundarySeq}
\begin{diagram}
0 \rightarrow \widetilde {\mathcal A} & \rTo{ \ \ i \ \ } & \widehat{\mathcal  A}  & \rTo{ \ \ \mathrm{ev} \ \ }  & \mathcal A \rightarrow 0,
\end{diagram}
\end{equation}
where $i$ is the inclusion map and $\mathrm{ev}$ the evaluation map such that
\begin{equation}
  \mathrm{ev} (H|_{\rm open}+\widetilde V_{\rm edge} ) = H.
\end{equation}
 Another exact sequence relates to the adiabatic cycle
 \begin{equation}
\label{Eq:SuspensionSeq}
\begin{diagram}
0 \rightarrow \mathcal S \mathcal A & \rTo{ \ \ i \ \ } & \mathcal A^{\mathbb S^1}  & \rTo{ \ \ \mathrm{ev} \ \ }  & \mathcal A \rightarrow 0,
\end{diagram}
\end{equation}
where $\mathcal S \mathcal A$ denotes the suspension of the algebra $\mathcal A$, that is, the ideal generated by the elements $H(\tau) \in \mathcal A^{\mathbb S^1}$ with $H(0)=0$. The evaluation map above is $\mathrm{ev} (H(\tau) )=H(0)$. Similar sequences exist for the suspensions of the algebras $\widehat{\mathcal A}$ and $\widetilde{ \mathcal A}$. They can all be combined in the following commutative diagram \cite{LeungJPA2020}:
\begin{diagram}
& & 0 & \  &  0 & \ & 0 & & \\
& & \uTo & \  &  \uTo & \ & \uTo & & \\
0 & \rTo  & \mathcal S \mathcal A & \rTo{ \ \ i \ \ } & \mathcal A^{\mathbb S^1}  & \rTo{\ \ \mathrm{ev} \ \ }  & \mathcal A & \rTo & 0 &
\\
& & \uTo & \  &  \uTo & \ & \uTo & & \\
0 & \rTo  & \mathcal S \widehat{\mathcal A} & \rTo{ \ \ i \ \ } & \widehat{\mathcal A}^{\mathbb S^1}  & \rTo{\ \ \mathrm{ev} \ \ }  & \widehat{\mathcal A} & \rTo & 0 &
\\
& & \uTo & \  &  \uTo & \ & \uTo & & \\
0 & \rTo  & \mathcal S \widetilde{\mathcal A} & \rTo{ \ \ i \ \ } & \widetilde{\mathcal A}^{\mathbb S_1}  & \rTo{\ \ \mathrm{ev} \ \ }  & \widetilde{\mathcal A} & \rTo & 0 & \\
& & \uTo & \  &  \uTo & \ & \uTo & & \\
& & 0 & \  &  0 & \ & 0 & & 
\end{diagram}
We point out that the bulk adiabatic process of one chain takes place inside the algebra located in the upper-left corner of the diagram. The relative index of projections mentioned in the main text is associated with the algebra of edge operators, which is located at the bottom-right corner of the diagram.

Every long exact sequence of $C^\ast$-algebras
\begin{equation}
\begin{diagram}
0 \rightarrow \mathcal J & \rTo{ \ \ i \ \ } & \mathcal B  & \rTo{ \ \mathrm{ev} \ \ \ \ }  & \mathcal B/\mathcal J \rightarrow 0
\end{diagram}
\end{equation}
induces a six-term exact sequence among the K-theories \cite{RLL}, which is central to the rigorous derivations of the bulk-boundary correspondences \cite{KRS, ProdanSpringer2016A}:
\begin{equation}\label{SixTermDiagramAlg}
\begin{diagram}
& K_0(\mathcal J) & \rTo{ \ \ i_\ast \ \ } & K_0(\mathcal B)  & \rTo{\ \ \mathrm{ev}_\ast \ \ } & K_0(\mathcal B/\mathcal J) &\\
& \uTo{\rm Ind}& \  &  \ & \ & \dTo{\rm Exp} & \\
& K_1(\mathcal B/\mathcal J)  & \lTo{\ \ \mathrm{ev}_\ast \ \ } & K_1(\mathcal B) & \lTo{\ \ i_\ast \ \ } & K_1(\mathcal J) &
\end{diagram}
\end{equation}
If the exact sequence is the particular sequence from Eq.~\ref{Eq:SuspensionSeq}, then the ${\rm Ind}$ and ${\rm Exp}$ maps are traditionally denoted by $\theta$ and $\beta$. These maps play a central role in Bott's periodicy theorem \cite{RLL}.

By applying this principle to the commuting diagram from above, we produce the following sequence
\begin{equation}\label{Eq:KDiagram}
\begin{diagram}
 & K_0(\mathcal S \mathcal A)   &  \rTo{\ \ \theta^{-1} \ \ } & K_1(\mathcal A) & \rTo{\rm Ind} & K_0(\widetilde{\mathcal A}),
\end{diagram}
\end{equation}
which will be our main tool for deriving the bulk-edge correspondence in Appendix~C. This diagram connects the complete topological invariants of the bulk adiabatic process of a single chain and of its edge physics. 

Let us point out that any projection $\tau \mapsto P(\tau)$ from $\mathcal A^{\mathbb S^1}$ generates a canonical element from $K_0(\mathcal S \mathcal A)$ via \begin{equation}
    [\tau \mapsto P(t)]_0 -[\tau \mapsto P(0)]_0.
\end{equation}
This will be tacitly assumed throughout. All the algebras introduced can be promoted to include weak disorder and all the statements made below remain valid for this more general setting (see Ref.~\cite{LeungJPA2020} for details).

\section{Bulk-edge correspondence and relative index}
\label{AppendixC}

Here we provide the proof to Eq.~\eqref{Eq:RelativeIndChern} and Eq.~\eqref{Eq:TopIndexMap} from the main text, which follows closely the proof of Prop~6.3 in Ref.~\cite{LeungJPA2020}. First, the inverse map $\theta^{-1}: K_0 (\mathcal S \mathcal A ) \to K_1 (\mathcal A)$ for the class of
\begin{equation}
    P(\tau)=\chi_{(-\infty, E_F]}\big( H(\gamma_{\tau}) \big), \quad \tau \in \mathbb S^1,
\end{equation}
in $K_0(\mathcal S \mathcal A)$ was computed in Proposition~4.1.8 of Ref.~\cite{ProdanSpringer2016A}. It is given by
\begin{equation}\label{Eq:thetaMap}
\theta^{-1} \big ( [\tau \mapsto P (\tau)]_0\big ) = [V_1 + 1 - P(0)]_1,
\end{equation}
where $V_{\tau}$ is the monodromy of $P(\tau)$, {\it i.e.} \begin{equation}\label{Eq:Monodromy}
V_\tau = U_\tau^A P(0),
\end{equation}
with $U_\tau^A$ the unitary solution of the adiabatic time evolution equation
\begin{equation}\label{Eq:TimeEvol2}
{\rm i} \partial_\tau U_{\tau}^A = \big (T \, H(\gamma_\tau)+i [ \partial_\tau P(\tau), P(\tau) ]\big ) U_{\tau}^A, \quad U^A_{0} = I.
\end{equation}
Then our next task is to compute 
\begin{equation}
 {\rm Ind}\big ( [V_1 + 1 - P(0)]_1 \big ),
\end{equation}
where ${\rm Ind}$ is the index map associated with the sequence~\eqref{Eq:BulkBoundarySeq}. 

The standard picture of the index map \cite{RLL} states that,  for any unitary operator $U \in \mathcal A$,
\begin{equation}\label{Eq:DefIndexMap}
{\rm Ind} \big([U]_1 \big) = \left [\widehat{\bm U} \begin{pmatrix} 1 & 0 \\ 0 & 0 \end{pmatrix} \widehat{\bm U}^\ast \right]_0 - \left [ \begin{pmatrix} 1 & 0 \\ 0 & 0 \end{pmatrix} \right ]_0,
\end{equation}
where $\widehat{\bm U}$ is a lift of the unitary operator ${\small \begin{pmatrix} U & 0 \\ 0 & U^\ast \end{pmatrix}}$
to the algebra $M_2 \otimes \widehat{\mathcal A}$, namely
\begin{equation}
\mathrm{ev} \big( \widehat{\bm U} \big) = \begin{pmatrix} U & 0 \\ 0 & U^\ast \end{pmatrix}.
\end{equation}
The key observation is that, if $U$ is the unitary $V_1+1- P (0)$, then $\widehat{\bm W}_1 + \bm 1 - \widehat{\bm Q}(0)$ from Eq.~\eqref{Eq:TheLift} supplies such a lift \cite{LeungJPA2020}. Indeed, if we apply the algebra homomorphism $\mathrm{ev}$ to Eq.~\eqref{Eq:TimeEvol1}, all edge operators are sent to zero and, as such, we obtain two copies of Eq.~\eqref{Eq:TimeEvol2}, of which one is ran backwards in time. Therefore, 
\begin{equation}\label{Eq:Ev1}
    \mathrm{ev}\big( \widehat {\bm W}_1 \big) =  \begin{pmatrix} V_1 & 0 \\ 0 & V_1^\ast \end{pmatrix}.
\end{equation} 
Furthermore, 
\begin{equation}
    \mathrm{ev} \big( \widehat{\bm Q}(0) \big) = \begin{pmatrix} P(0) & 0 \\ 0 & P(0) \end{pmatrix}.
\end{equation}
Then
\begin{align}
& \qquad \big ({\rm Ind}\circ \theta^{-1}\big ) \big ([\tau \mapsto P (\tau) ]_0 \big) = \\ \nonumber
&\left [\big (\widehat{\bm W}_{1}+ {\bm 1} -\widehat{\bm Q}(0)\big ) {\small \begin{pmatrix} 1 & 0 \\ 0 & 0 \end{pmatrix}} \big (\widehat{\bm W}_{1}+ {\bm 1}-\widehat{\bm Q}(0)\big )^\ast \right]_0 - \left [ {\small \begin{pmatrix} 1 & 0 \\ 0 & 0 \end{pmatrix}} \right ]_0.
\end{align}
Since we assume that $\widetilde{\bm V}_{\rm edge}(0)=0$, the following simplification occurs:
\begin{equation}
\widehat{\bm Q}(0) = \begin{pmatrix} \widehat{P}(0) & 0 \\ 0 & \widehat{P}(0) \end{pmatrix}
\end{equation}
Hence, ${\bm 1}-\widehat{\bm Q}(0)$ commutes with ${\small \begin{pmatrix} 1 & 0 \\ 0 & 0 \end{pmatrix}}$. Additionally, we have the relation $\widehat{\bm W}_{1} ( {\bm 1} -\widehat{\bm Q}(0) )=0$. We reach the conclusion
\begin{equation}\label{Eq:Ind}
    \big ({\rm Ind}\circ \theta^{-1} \big) \big ( [\tau \mapsto P (\tau)]_0 \big) = \big [ \widehat{\bm \Pi}^A_{1} \big]_0 \ - \ \big [ \widehat{\bm \Pi}^{}_0 \big ]_0,
\end{equation}
where \begin{equation}
\widehat{\bm \Pi}^A_{1}=\widehat{\bm U}_{1}^A \begin{pmatrix} \widehat{P}(0) & 0 \\ 0 & 0 \end{pmatrix} \widehat{\bm U}_{1}^{A\ast} \ \ \mbox{and} \ \ \widehat{\bm \Pi}^{}_0 =\begin{pmatrix} \widehat{P}(0) & 0 \\ 0 & 0 \end{pmatrix},
\end{equation}
as claimed in the main text.

Regarding the relative index of the projections, we first note that, due to Eq.~\eqref{Eq:Ev1},
\begin{equation}
    \mathrm{ev}\big( \widehat{\bm \Pi}^A_{1}-\widehat{\bm \Pi}^{}_{0} \big) =0,
\end{equation}
which assures us that $\widehat{\bm \Pi}^A_{1}-\widehat{\bm \Pi}^{}_{0} \in \widetilde{\mathcal A} \simeq M_\infty$ (this is a special feature of one-dimensional systems). Then the spectrum of the self-adjoint operator $\widehat{\bm \Pi}^A_{1}-\widehat{\bm \Pi}^{}_{0} $ is discrete (except at the origin) and the possible eigenvalues $\pm 1$ have finite degeneracies. Furthermore, there exists a unitary element $K$ from the algebra $\mathbb C \cdot I \oplus M_\infty$ such that 
\begin{equation}\label{Eq:K}
 K \; \widehat{\bm \Pi}^A_{1}\;  K^\ast-\widehat{\bm \Pi}^{}_{0} = P_{+1}-P_{-1},
\end{equation}
where $P_{\pm 1}$ are the spectral projections of $\widehat{\bm \Pi}^A_{1}-\widehat{\bm \Pi}^{}_{0} $ onto $\pm 1$. The relative index is then \cite{AvronPRL1990}
\begin{equation}
    {\rm Index}\big (\widehat{\bm \Pi}^A_{1},\ \widehat{\bm \Pi}^{}_{0} \big ) = {\rm rank } \, P_{+1}- {\rm \, rank}\, P_{-1}.
\end{equation}
Eq.~\eqref{Eq:K} also assures us that 
\begin{equation}
\big [ \widehat{\bm \Pi}^A_{1} \big]_0 \ - \ \big [ \widehat{\bm \Pi}^{}_0 \big ]_0=\big [ P_{+1} \big]_0 \ - \ \big [ P_{-1} \big ]_0,
\end{equation}
as an element of $K_0(\widetilde{\mathcal A})$, hence Eq.~\eqref{Eq:Ind} can be written as
\begin{equation}
    \big ({\rm Ind}\circ \theta^{-1} \big) \big ( [\tau \mapsto P (\tau)]_0 \big) = \big [ P_{+1} \big]_0 \ - \ \big [ P_{-1}\big ]_0.
\end{equation}
By applying the standard trace, on the right side we get the relative index, while on the left side we get the Chern number of $[\tau \mapsto P (\tau)]$. The latter follows from the behavior of the cyclic cocycle pairings relative to the connecting maps, worked out in details in Ref.~\cite{ProdanSpringer2016A}[Sec.~5.5].


\begin{thebibliography}{10}

\bibitem{Thouless1983} D. J. Thouless, {\sl Quantization of particle transport}, Phys. Rev. B {\bf 27}, 6083 -6087 (1983).

\bibitem{AltshulerScience1999} B. L. Altshuler and L. I. Glazman, {\sl Pumping electrons}, Science {\bf 283}, 1864-1865 (1999).

\bibitem{RiceMele1982} M. J. Rice and E. J. Mele, {\sl Elementary excitations of a linearly conjugated diatomic polymer}, Phys. Rev. Lett., {\bf 49}, 1455-1459 (1982).

\bibitem{NiuRMP2010} D. Xiao, M.-C. Chang, and Q. Niu, {\sl Berry phase effects on electronic properties}, Rev. Mod. Phys. {\bf 82}, 1959 (2010).

\bibitem{KRS} J. Kellendonk, T. Richter, and H. Schulz-Baldes, {\sl  Edge current channels and Chern numbers in the integer quantum Hall effect}, Rev. Math. Phys. {\bf 14}, 87-119 (2002).

\bibitem{ProdanSpringer2016A} E. Prodan and H. Schulz-Baldes, {\sl Bulk and Boundary Invariants for Complex Topological Insulators: From K-theory to Physics}, (Springer, Berlin, 2016).

\bibitem{LeungJPA2020} B. Leung and E. Prodan, {\sl Bulk-boundary correspondence for topological insulators with quantized magneto-electric effect}, J. Phys. A: Math. Theor. {\bf 53}, 205203 (2020).

%cold atoms

\bibitem{NakajimaNature2016} S. Nakajima, T. Tomita, S. Taie, T. Ichinose, H. Ozawa, L. Wang, M. Troyer, and Y. Takahashi, {\sl Topological Thouless pumping of ultracold fermions}, Nature Phys, {\bf 12}, 296–300 (2016).

\bibitem{LohseNature2016} M. Lohse, C. Schweizer, O. Zilberberg, M. Aidelsburger, and I. Bloch, {\sl A Thouless quantum pump withultracold bosonic atoms in an optical superlattice}, Nature Phys. {\bf 12}, 350–354 (2016).

%edge-to-edge pumping

\bibitem{GrinbergNatComm2020} I. H. Grinberg, M. Lin, C. Harris, W. A. Benalcazar, C. W. Peterson, T. L. Hughes, and G. Bahl, {\sl Robust temporal pumping in a magneto-mechanical topological insulator}, Nat. Commun. {\bf 11}, 974 (2020).

\bibitem{XiaPRL2021} Y. Xia, E. Riva, M. I. N. Rosa, G. Cazzulani, A. Erturk, F. Braghin, and M. Ruzzene, {\sl Experimental observation of
temporal pumping in electro-mechanical waveguides}, Phys. Rev. Lett. {\bf 126}, 095501 (2021).

\bibitem{ChengPRL2020} W. Cheng, E. Prodan, and C. Prodan, {\sl Experimental demonstration of dynamic
topological pumping across incommensurate bilayered acoustic metamaterials}, Phys. Rev. Lett. {\bf 125}, 224301 (2020).

\bibitem{XuPRL2020} X. Xu, Q. Wu, H. Chen, H. Nassar, Y. Chen, A. Norris, M. R. Haberman, and G. Huang, {\sl Physical observation of a robust acoustic pumping in waveguides
with dynamic boundary}, Phys. Rev. Lett. {\bf 125}, 253901 (2020).

% spin-Chern insulator

\bibitem{LvPRL2021} Q.-X. Lv, Y.-X. Du, Z.-T. Liang, H.-Z. Liu, J.-H. Liang, L.-Q. Chen, L.-M. Zhou, S.-C. Zhang, D.-W. Zhang, B.-Q. Ai, H. Yan, and S.-L. Zhu, {\sl Measurement of spin Chern numbers in quantum simulated topological insulators}, Phys. Rev. Lett. {\bf 127}, 136802 (2021).

% Adiabatic theorem

\bibitem{Kato} T. Kato, {\sl On the adiabatic theorem of quantum mechanics}, J. Phys. Soc. Japan {\bf 5}, 435-439 (1950).

\bibitem{AvronCMP1987} J. E. Avron, R. Seiler, and L. G. Yaffe, {\sl Adiabatic Theorems and Applications to the Quantum Hall Effect}, Commun. Math. Phys. {\bf 110}, 33-49 (1987). 


% relative index of projections

\bibitem{AvronPRL1990} J. Avron, R. Seiler, and B. Simon, {\sl The Quantum Hall Effect and the Relative Index for Projections}, Phys. Rev. Lett. {\bf 65}, 2185-2188 (1990).

\bibitem{AvronCMP1994}  J. Avron, R. Seiler, and B. Simon, {\sl Charge deficiency, charge transport and comparison of dimensions}, Commun. Math. Phys. {\bf 159}, 399-422 (1994).

\bibitem{AvronJFA1994}  J. Avron, R. Seiler, and B. Simon, {\sl The index of a pair of projections}, J. Funct. Anal. {\bf 120}, 220-237 (1994). 

% Chern number

\bibitem{BellissardJMP1994} J. Bellisard, A. van Elst, and H. Schulz-Baldes, {\sl The noncomutative geometry of the quantum Hall effect}, J. Math. Phys. {\bf 35}, 5373–451 (1994).

\bibitem{Laughlin} R. B. Laughlin, {\sl Quantized Hall conductivity in two dimensions}, Phys. Rev. B {\bf 23}, 5632 (1981). 

\bibitem{Footnote1} A topological invariant is called complete if it determines the classification entirely.

\bibitem{Supplemental} See Supplemental Material which also contains Refs.~\cite{LeungJPA2020, ProdanSpringer2016A, KRS, Kato, AvronCMP1987, AvronPRL1990, RLL}. Appendix~A contains the adiabatic theorem and defines the adiabatic time evolution. Appendix~B describes the algebras of our physical operators and their relations via exact sequences. Appendix~C details the K-theoretic calculations to derive the bulk-edge correspondence for our system and its connection to the relative index. The calculations prove our main result.

\bibitem{RLL} M. Rordam, F. Larsen, and N. Laustsen, {\sl An Introduction to K-Theory for $C^\ast$-Algebras}, (Cambridge University, Cambridge, England, 2000).

\end{thebibliography}
\end{document}